\documentclass[aps,pra,groupedaddress,nofootinbib,notitlepage,showpacs,floatfix,twocolumn]{revtex4-1}
\usepackage{graphicx}
\usepackage{epstopdf}
\usepackage{amsmath}
\usepackage{amssymb}
\usepackage{hyperref}
\usepackage{bm}
\usepackage{subfigure}

\hypersetup{
    colorlinks=true,       % false: boxed links; true: colored links
    linkcolor=cyan,          % color of internal links
    citecolor=magenta,        % color of links to bibliography
    filecolor=magenta,      % color of file links
    urlcolor=cyan,           % color of external links
    runcolor=cyan
}

\newcommand{\beq}{\begin{equation}}
\newcommand{\eeq}{\end{equation}}
\newcommand{\beqnn}{\begin{equation*}}
\newcommand{\eeqnn}{\end{equation*}}
\newcommand{\bea}{\begin{eqnarray}}
\newcommand{\eea}{\end{eqnarray}}
\newcommand{\beann}{\begin{eqnarray*}}
\newcommand{\eeann}{\end{eqnarray*}}
\newcommand{\bes}{\begin{subequations}}
\newcommand{\ees}{\end{subequations}}
\newcommand{\braket}[2]{\langle #1 | #2\rangle}
\newcommand{\ket}[1]{ | #1\rangle}
\newcommand{\bra}[1]{\langle #1 | }
\newcommand{\ketbra}[2]{|#1\rangle\langle #2|}
\newcommand{\ident}{\openone}

\newcommand{\ignore}[1]{}

\begin{document}

\title{Coherent control of non-Markovian photon resonator dynamics}

\author{A. F. J. Levi}

\affiliation{Department of Physics and Astronomy, University of Southern California,
Los Angeles, CA 90089-0484}

\affiliation{Department of Electrical Engineering, University of Southern California,
Los Angeles, CA 90089-2533}

\author{L. Campos Venuti}

\affiliation{Department of Physics and Astronomy, University of Southern California,
Los Angeles, CA 90089-0484}

\author{T. Albash}

\affiliation{Department of Physics and Astronomy, University of Southern California,
Los Angeles, CA 90089-0484}

\author{S. Haas}

\affiliation{Department of Physics and Astronomy, University of Southern California,
Los Angeles, CA 90089-0484}

\date{\today}
\begin{abstract}
We study the unitary time evolution of photons interacting with a
dielectric resonator using coherent control pulses. We show that non-Markovianity
of transient photon dynamics in the resonator subsystem may be controlled
to within a photon-resonator transit time. In general, appropriate
use of coherent pulses and choice of spatial subregion may be used
to create and control a wide range of non-Markovian transient dynamics
in photon resonator systems. 
\end{abstract}


\maketitle

\section{Introduction}

\label{sec1}

The transient dynamics of photons interacting with a resonator is
of both fundamental and practical interest. For example, microcavity
resonators have been explored as a means to delay light in classical
communication systems~\cite{Yariv}, and similar ideas have been developed
for single photons~\cite{Preble,Kimble,Fan} with potential future
use in quantum communication protocols. These and other studies exploit
a basic property of a resonator subsystem, namely the ability to store
photon energy density and release it at a later time. Since Markovian
dynamics may be identified with information flow leaving the system~\cite{Breuer},
it seems natural to expect that storage of a photon or many photons
in a resonator can result in non-Markovian behavior. More precisely,
if we consider a finite region of a resonator, energy can both enter
and leak out of the designated region depending on the interplay of
system parameters and the location of the region itself. It seems
natural to expect a high degree of non-Markovianity in such a situation.
Furthermore, non-Markovianity may be viewed as a resource for information
processing tasks~\cite{Maniscalco}. One is therefore motivated to
demonstrate control of photon transient dynamics and hence control
of the associated non-Markovianity.

\begin{figure}[ht]
\includegraphics[width=0.6\columnwidth]{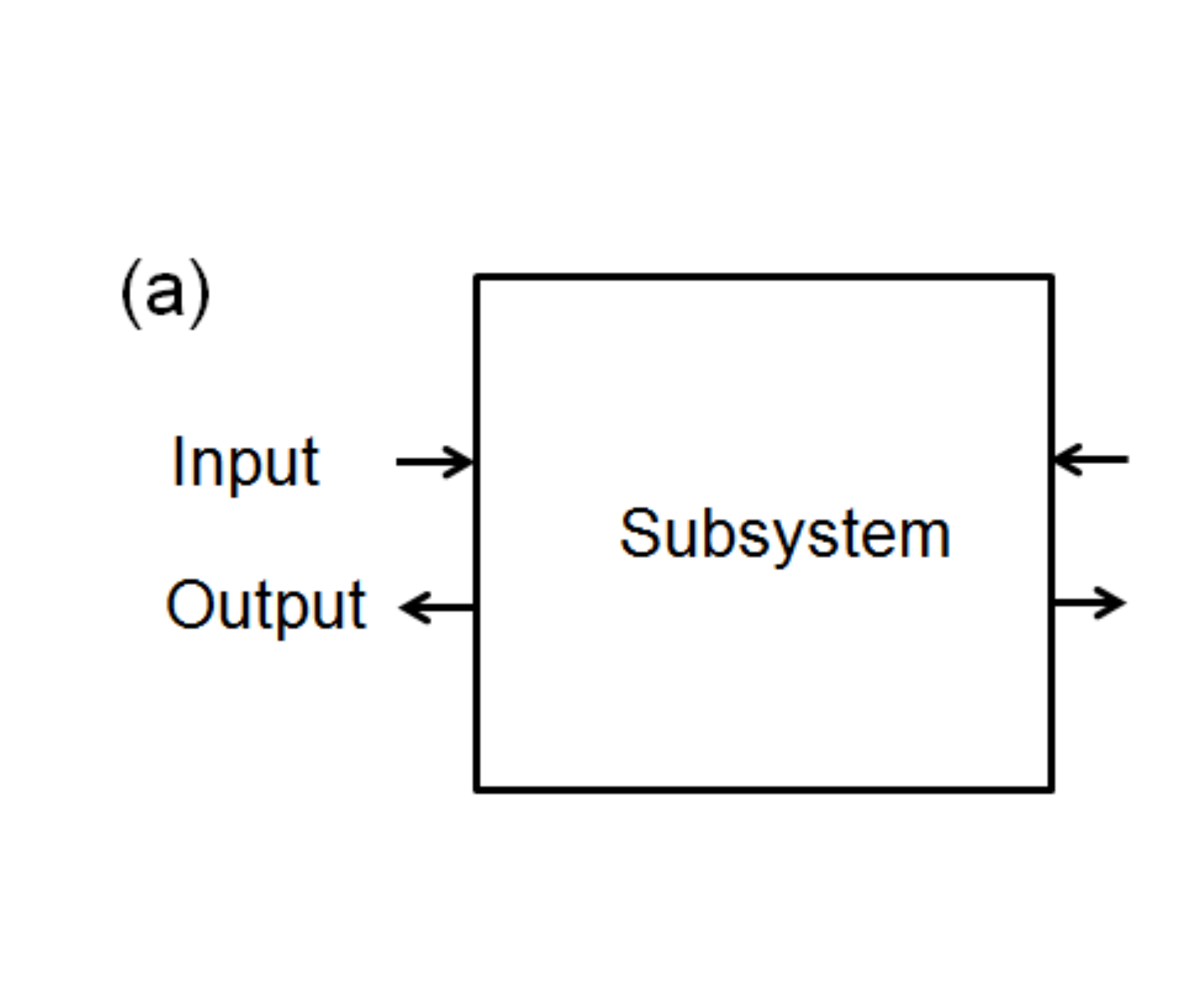} 
\includegraphics[width=0.6\columnwidth]{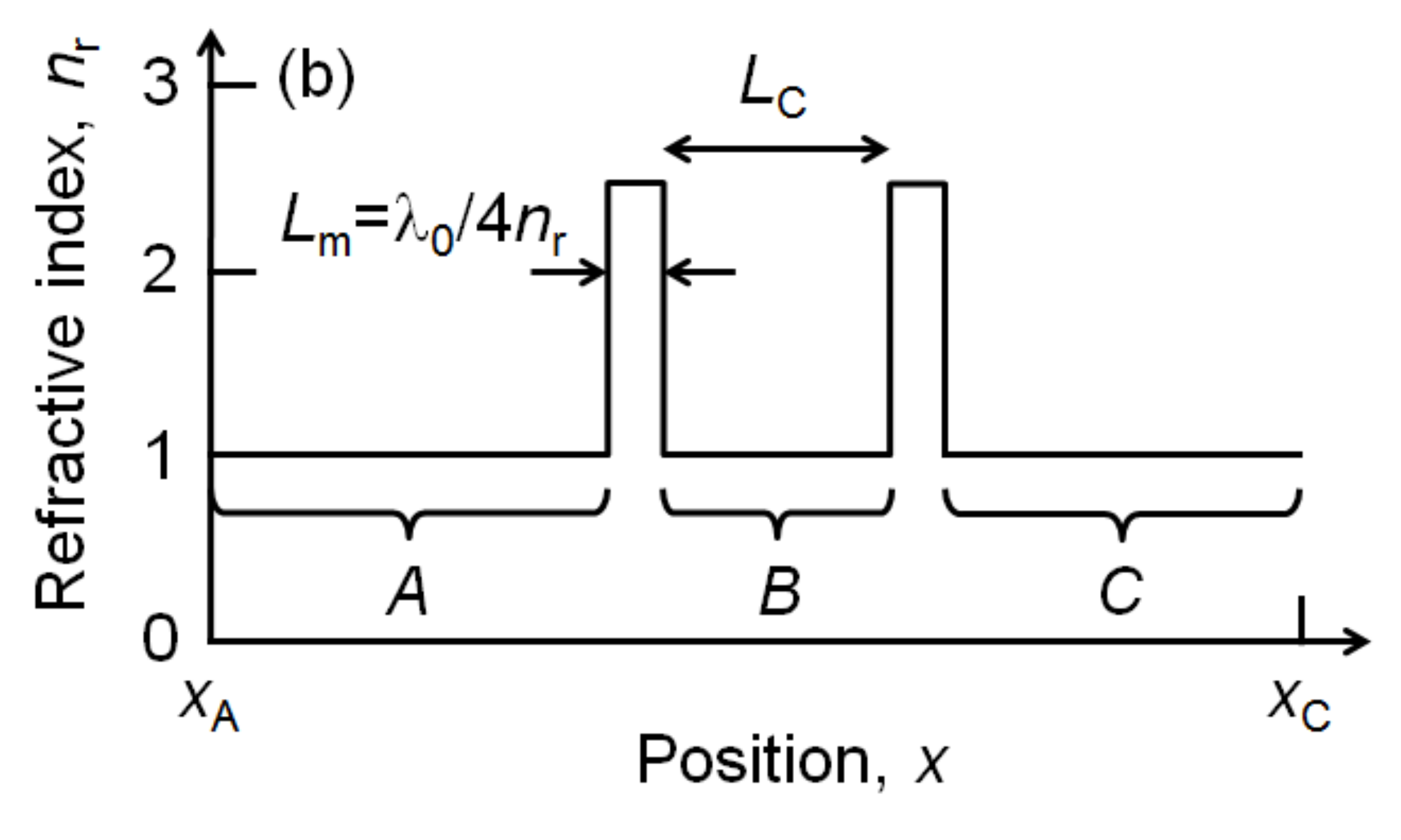} \caption{\label{fig1} {\footnotesize { (a) Sketch of a subsystem with inputs
and outputs to continuum states. (b) Symmetric Fabry-P/'erot resonator
subsystem divided into spatial regions $A$, $B$, and $C$ separated
by a quarter-wavelength lossless dielectric characterized by refractive
index $n_{{\rm {r}}}$. The resonant wavelength is $\lambda_{0}$.}}}
\end{figure}

To investigate such non-Markovian effects we study the full time evolution
of a Hamiltonian system and concentrate on the dynamics of a subregion
obtained by tracing out exactly the remaining degrees of freedom.
In a unitary system of finite spatial extent, excitations are reflected
indefinitely back and forth from the boundaries, and consequently
any subregion of such a system would always display non-Markovian
behavior. The same holds true for a system with discrete energy levels
because of the formation of bound states. To avoid these trivial cases,
we seek therefore a system with a continuous energy spectrum such
that the subsystem can exchange continuous energy with its environment
as sketched in Fig.~\ref{fig1}(a).

The physics we are interested in exploring may be captured by a single
resonator with a refractive index profile as illustrated in Fig.~\ref{fig1}(b).
The symmetric one-dimensional Fabry-P/'erot resonator consists of three
spatial regions, $A$, $B$, and $C$ in vacuum separated by two lossless
dielectric mirrors, each of refractive index $n_{{\rm {r}}}$ and
thickness $L_{{\rm {m}}}=\lambda_{0}/4n_{{\rm {r}}}$, where $\lambda_{0}$
is the resonant photon wavelength in vacuum. Spatial regions $A$
and $C$ connect to continuous input and output states at $x=x_{{\rm {A}}}$
and $x=x_{{\rm {C}}}$, respectively. The resonator cavity length is
$L_{{\rm {B}}}$ and defines the spatial extent of region $B$. At
the resonant photon wavelength, the complex mirror reflection amplitude
is $re^{i\pi}=-r$, and the transmission amplitude is $te^{i\pi/2}=it$.
Flux conservation in the lossless system requires $|r|^{2}+|t|^{2}=1$.
Transmission through each mirror depends weakly on wavelength such
that 
\begin{equation}
|t|^{2}=\frac{1}{1+\left(\frac{k_{1}^{2}-k_{2}^{2}}{2k_{1}k_{2}}\right)^{2}\sin^{2}(k_{2}L_{{\rm {m}}})}\ ,\label{eq1}
\end{equation}
 where the propagation constant in vacuum is $k_{1}=2\pi/\lambda$
and in the dielectric mirror it is $k_{2}=2\pi n_{{\rm {r}}}/\lambda$.

A single photon may be described by a wave function $\Psi(x,t)$ with
the interpretation that $|\Psi(x,t)|^{2}$ is the photon energy density~\cite{Birula,Sipe,Smith}.
We choose to use a single-photon wave function description because,
as will become apparent, it has the significant advantages in this
initial study of both simplicity and ease of interpretation. The unitary
dynamics of the photon wave function propagating in the $x$ direction
in a lossless dielectric media may be modeled as a phase-coherent
integral of linearly polarized basis states $\phi_{\omega}(x)$ with
amplitudes $\alpha_{\omega}$ ,
\begin{equation}
\Psi(x,t)=\int\frac{d\omega}{2\pi}\alpha_{\omega}\phi_{\omega}(x)e^{-i\omega t}\ ,\label{eq2}
\end{equation}
 where, as shown in the Appendix, $\phi_{\omega}(x)$ is a normalized
solution of the one-dimensional Helmholtz equation,
\begin{equation}
\frac{d}{dx}\left(\frac{1}{\mu_{\mathrm{r}}(x)}\frac{d}{dx}\phi_{\omega}(x)\right)+\omega^{2}\epsilon_{\mathrm{r}}(x)\epsilon_{0}\mu_{0}\phi_{\omega}(x)=0\ .\label{eq3}
\end{equation}

The permeability of vacuum $\mu_{0}$ and permittivity of vacuum $\epsilon_{0}$
are related to the speed of light in vacuum via $c=1/\sqrt{\epsilon_{0}\mu_{0}}$.
Assuming a lossless dielectric material, the spatial profile may be characterized
by piecewise-constant values of relative permeability $\mu_{{\rm {r}}}$
and relative permittivity $\epsilon_{{\rm {r}}}$ in each region of
the domain; the conditions imposed on $\phi_{\omega}(x)$ at the boundary
between regions 1 and 2 at position $x_{0}$ are
\begin{equation}
\left.\phi_{\omega}(x)\right|_{x=x_{0}-\delta}=\left.\phi_{\omega}(x)\right|_{x=x_{0}+\delta}\ ,\label{eq4}
\end{equation}
and
\begin{equation}
\left.\frac{1}{\mu_{{\rm {r_{1}}}}}\frac{\mathrm{d}}{\mathrm{d}x}\phi_{\omega}(x)\right|_{x=x_{0}-\delta}=\left.\frac{1}{\mu_{{\rm {r_{2}}}}}\frac{\mathrm{d}}{\mathrm{d}x}\phi_{\omega}(x)\right|_{x=x_{0}+\delta}.\label{eq5}
\end{equation}

The refractive index is $n_{{\rm {r}}}=\sqrt{\mu_{{\rm {r}}}}\sqrt{\epsilon_{{\rm {r}}}}$.
If we assume that the photon coherence time is longer than any other
characteristic time scale, we may simply solve Eq.~(\ref{eq3}) to
completely describe the evolution of the photon. In the thermodynamic
limit there are a large number of photons in the system and Eq.~(\ref{eq3})
may also be used with the interpretation that the wave function corresponds
to the classical electric field~\cite{Birula,Sipe,Smith}. This means
that our model simultaneously describes a single photon and a classical
electromagnetic field.

An efficient and accurate way to solve Eq.~(\ref{eq3}) for the Fabry-P/'erot
resonator subsystem coupled to continuous input and output states
is to use the propagation matrix method~\cite{Kane,Levi}. All numerical
simulations we present as part of our study use this method.

\section{Transient response}

\label{sec2} We consider the transient response of a rectangular
single-photon pulse traveling left-to-right and incident on the Fabry-P/'erot
resonator. We smoothen the rectangular pulse with center frequency
$\omega_{0}$ by modulating a sinc function by a cosine in order to
reduce the Gibbs phenomenon. In this way a rectangular pulse of duration
$2T_{0}$ (length $2T_{0} c$) with rise and fall time $\tau_{{\rm {r}}}=2\pi/\Delta\omega_{{\rm {r}}}$
may be written as
\begin{eqnarray}
\Psi(x,t)&=&\int\displaylimits_{\left|\omega-\omega_{0}\right|\le\Delta\omega_{\mathrm{r}}}\frac{d\omega}{2\pi}\left(1+\cos\left(\frac{\pi(\omega-\omega_{0})}{\Delta\omega_{{\rm {r}}}}\right)\right) \nonumber \\
&& \times \frac{\sin((\omega-\omega_{0})T_{0})}{(\omega-\omega_{0})T_{0}}\phi_{\omega}(x)e^{-i\omega t}\ .\label{eq6}
\end{eqnarray}

To connect to existing photon technology we choose a cavity with resonant
wavelength $\lambda_{0}=1500\ {\rm {nm}}$ and resonant frequency
$\omega_{0}=2\pi/\tau_{0}$, where $\tau_{0}=5\ {\rm {fs}}$ corresponds
to a resonant photon energy of $E_{0}=\hbar\omega_{0}=0.827\ {\rm {eV}}$.
The refractive index of the mirrors is chosen to be $n_{{\rm {r}}}=2.5$,
region $A$ has length $L_{{\rm {A}}}$, region $C$ has length $L_{{\rm {C}}}$
and, unless stated otherwise, the resonator cavity length is $L_{{\rm {B}}}=15\times\lambda_{0}$.
The photon cavity round-trip time is $\tau_{{\rm {RT}}}=2L_{{\rm {B}}}/c=2\pi/\Delta\omega=30\times\tau_{0}=150\ {\rm {fs}}$,
and the resonator quality factor is $Q=144$, where $\tau_{Q}=Q/\omega_{0}=114\ {\rm {fs}}$.
Typically, one describes a transient response dominated by the ring-down
time constant $\tau_{Q}=1/\Gamma$, where the photon energy density of
a loaded resonator decays as $e^{-t/\tau_{Q}}$ and in which $\tau_{Q}$
is connected via a Fourier transform to a steady-state Lorentzian
energy density spectrum~\cite{Born,Saleh} ,
\begin{equation}
S(\omega)=\frac{S_{0}}{(\omega-\omega_{0})^{2}+(\Gamma/2)^{2}}\ .\label{eq7}
\end{equation}
 However, the actual transient dynamics of the system we wish to control
is more complex than this description would suggest.

\begin{figure}[ht]
\begin{centering}
\vspace{0cm}
 \includegraphics[width=0.7\columnwidth]{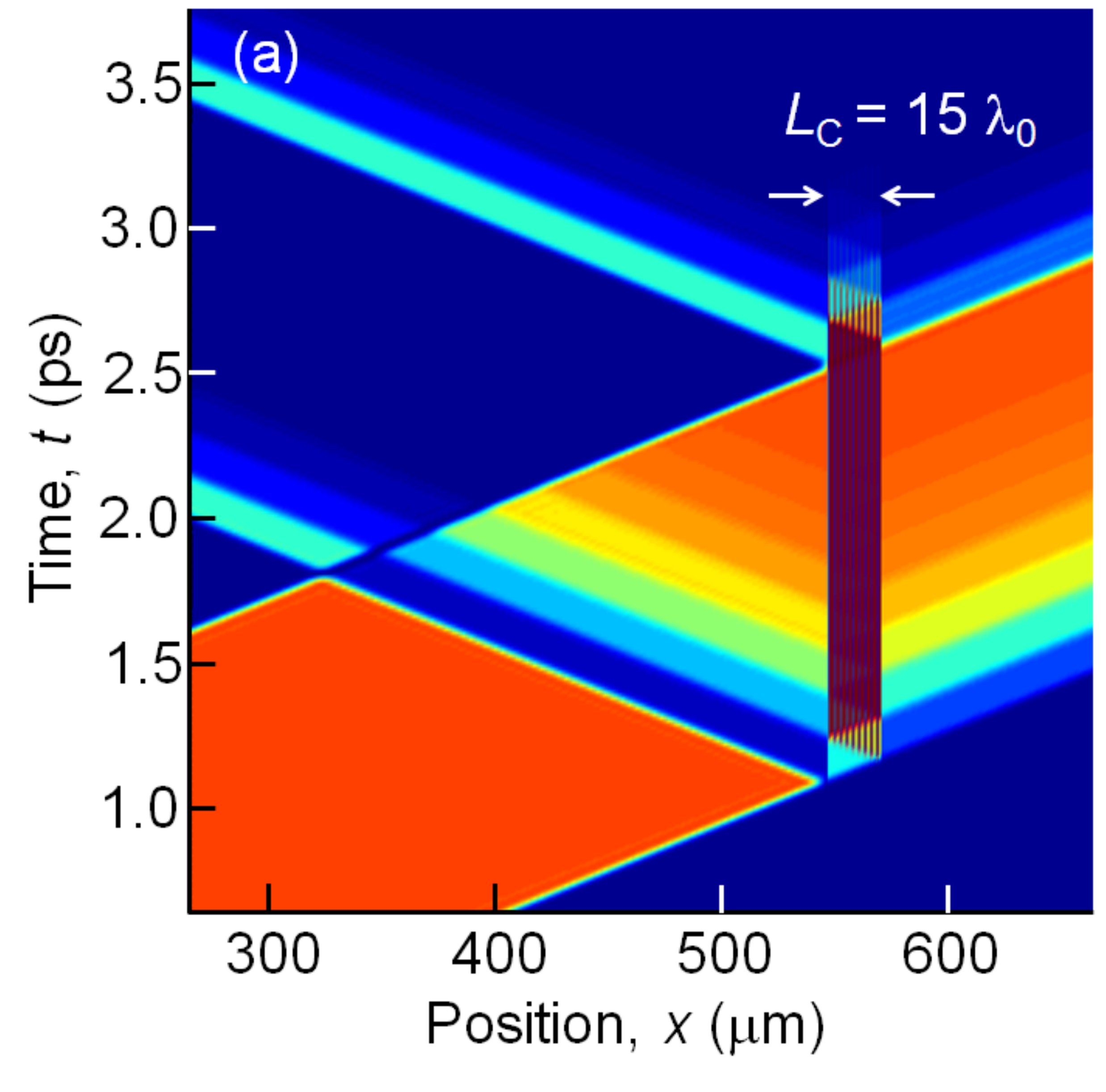} 
\par\end{centering}
\centering{}\includegraphics[width=0.7\columnwidth]{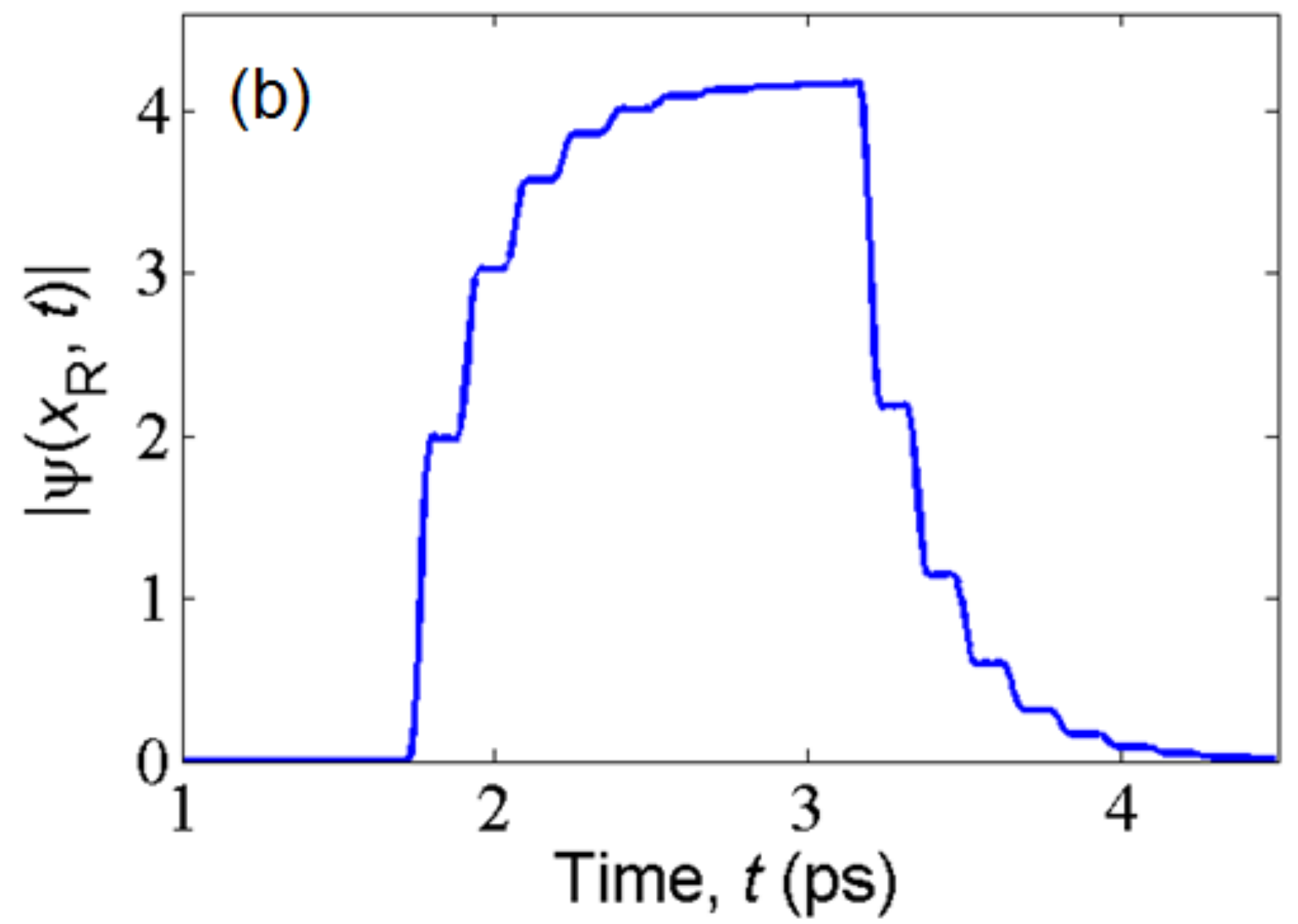} \caption{\label{fig2}(color online) {\footnotesize { (a) Space-time photon
energy density plot of a rectangular pulse incident on a Fabry-P/'erot
resonator. The resonator is of length $L_{B}$ (indicated by two arrows).
(b) $|\psi(x_{R},t)|$ (arbitrary scale) as a function of time detected
at position $x_{R}$ far to the right of the resonator. The field decay
constant $2\tau_{{\rm {Q}}}=229\ {\rm {fs}}$ is modulated by stepwise
response at the resonant cavity round-trip time $\tau_{{\rm {RT}}}=30\tau_{0}=150\ {\rm {fs}}$.
Photon pulse parameters are $\hbar\omega_{0}=0.827\ {\rm {eV}}$,
$\hbar\Delta\omega_{{\rm {s}}}=0.207\ {\rm {eV}}$, and $T_{0}\omega_{0}=900$.}}}
\end{figure}

Figure~\ref{fig2}(a) shows the calculated space-time photon energy
density plot of a rectangular pulse initially moving left to right
and incident on the Fabry-P/'erot resonator. The presence of the resonator
imparts temporal structure onto reflected and transmitted photon energy
density. The reflection at the leading edge and trailing edge of the
incident pulse is due to frequency components associated with the
pulse transient rise and fall times and the changing energy density
in the resonator. Subsequent reflections decay temporally in a stepwise
fashion in time steps of duration $\tau_{{\rm {RT}}}$. Figure~\ref{fig2}(b)
shows $|\Psi(x_{R},t)|$ calculated as a function of time detected
at position $x_{R}$ far to the right of the resonator. Photon energy
density both in the resonator and transmitted to position $x_{R}$
does not increase (or decay) as a simple exponential; rather, there
is a stepwise buildup (or decay) at each resonant cavity photon round-trip
time, $\tau_{{\rm {RT}}}$~\cite{OGorman}. With increasing rectangular
pulse duration, energy density asymptotically approaches the steady-state
value, which, on resonance at frequency $\omega_{0}$, results in
unity transmission and maximum energy density in the resonator. However,
our interest is not the steady state; rather, we seek to coherently
control the transient photon-resonator interaction using interference
effects and in this way control non-Markovianity of the system. The
shortest timescale on which we seek to exert control is the photon
cavity transit time $\tau_{{\rm {RT}}}/2$.

\begin{figure}[h]
\begin{centering}
\includegraphics[width=6.3cm]{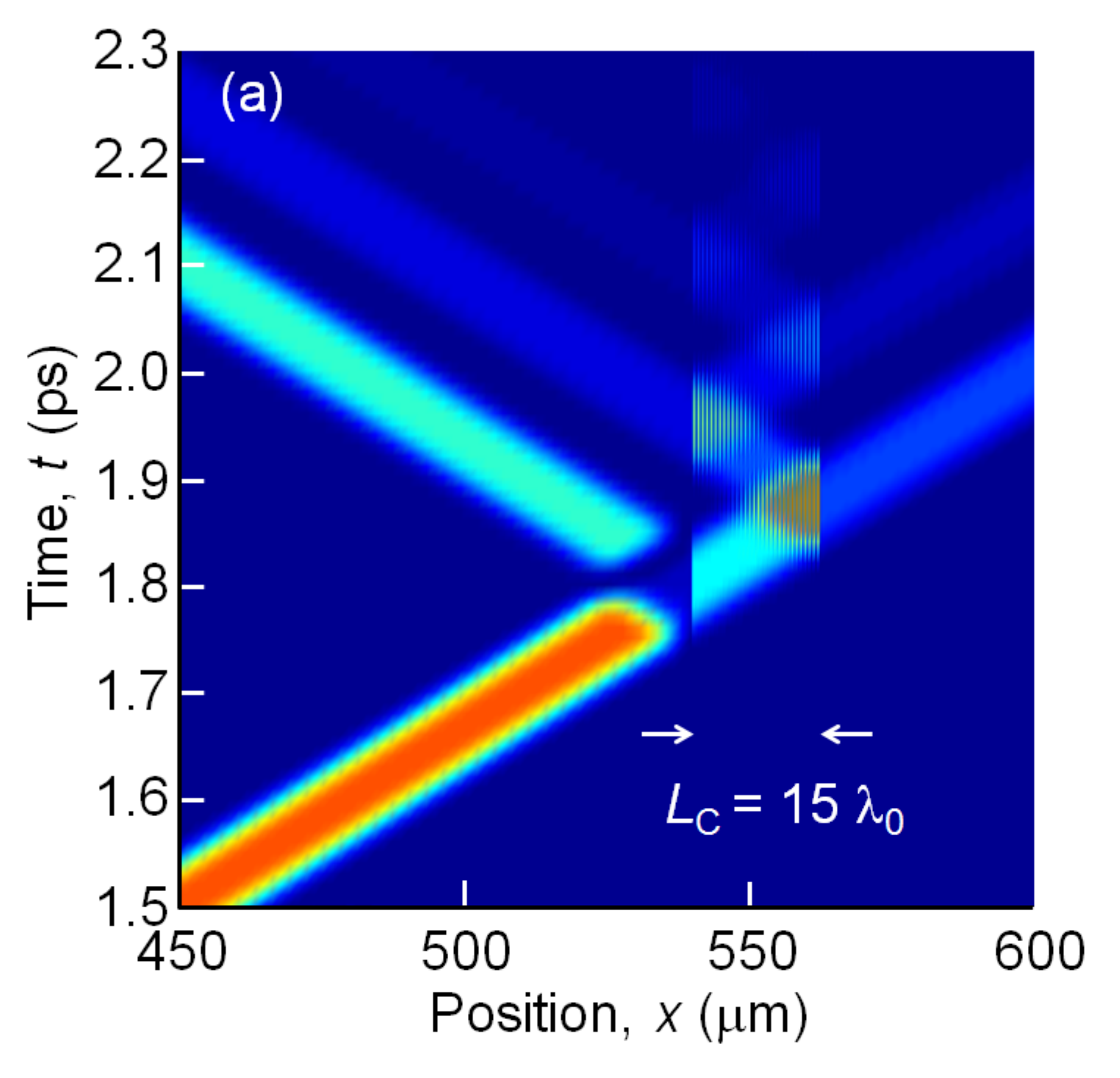} 
\par\end{centering}

\centering{}\includegraphics[width=7.3cm]{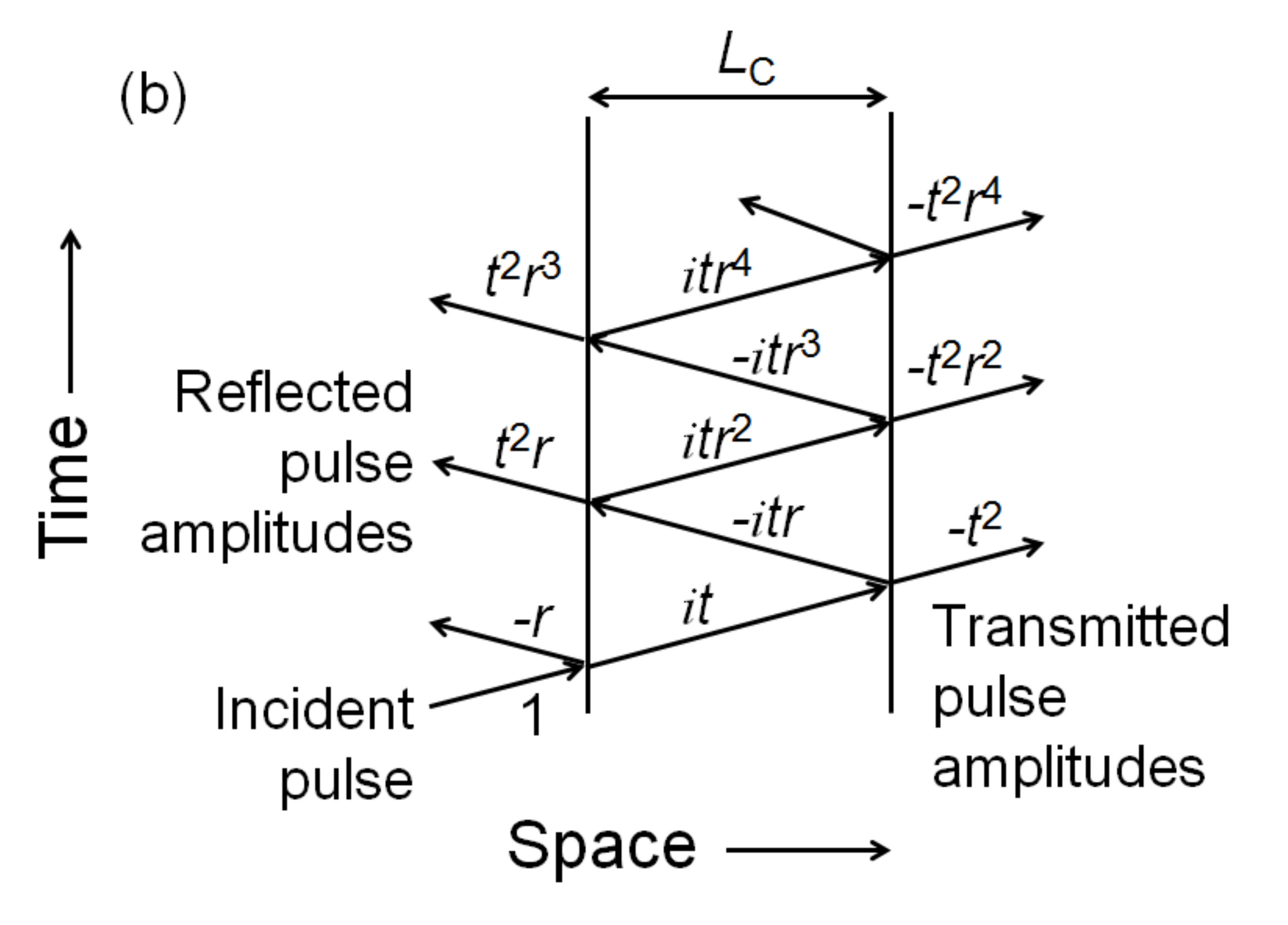} \caption{\label{fig3} {\footnotesize { (a) Space-time photon
energy density plot of a short rectangular pulse incident on the Fabry-P/'erot
resonator showing ring-down. (b) Space-time resonant photon ray trace
illustrating ring-down in the form of multiple transmitted and reflected
amplitudes. Photon pulse parameters are $\hbar\omega_{0}=0.827\ {\rm {eV}}$,
$\hbar\Delta\omega_{{\rm {s}}}=0.207\ {\rm {eV}}$, and $T_{0}\omega_{0}=60$.}}}
\end{figure}

Physical intuition and development of control concepts are best illustrated
using a photon pulse whose duration is short compared to the cavity
round-trip time, i.e., $2T_{0}<\tau_{{\rm {RT}}}$. Figure~\ref{fig3}(a)
shows the space-time photon energy density plot of a short rectangular
pulse initially moving left-to-right and incident on the Fabry-P/'erot
resonator. Initially, the photon energy density pulse entering the
cavity shows no indication of wave character. It is only after reflection
from the right mirror that self-interference effects are observed
and photon resonances inside the cavity begin to build up. The energy
stored in the resonator leaks out as forward and backscattered pulses.
The shortest time between forward and backscattered pulses is the
photon cavity transit time $\tau_{{\rm {RT}}}/2$.

Figure~\ref{fig3}(b) illustrates the origin of the ring-down observed
in the space-time diagram using space-time \emph{resonant} photon
ray tracing of reflected and transmitted amplitudes. The scattered
amplitudes at each mirror form a geometric series.

\begin{figure}[h]
\begin{centering}
\vspace{0cm}
 \includegraphics[width=6.3cm]{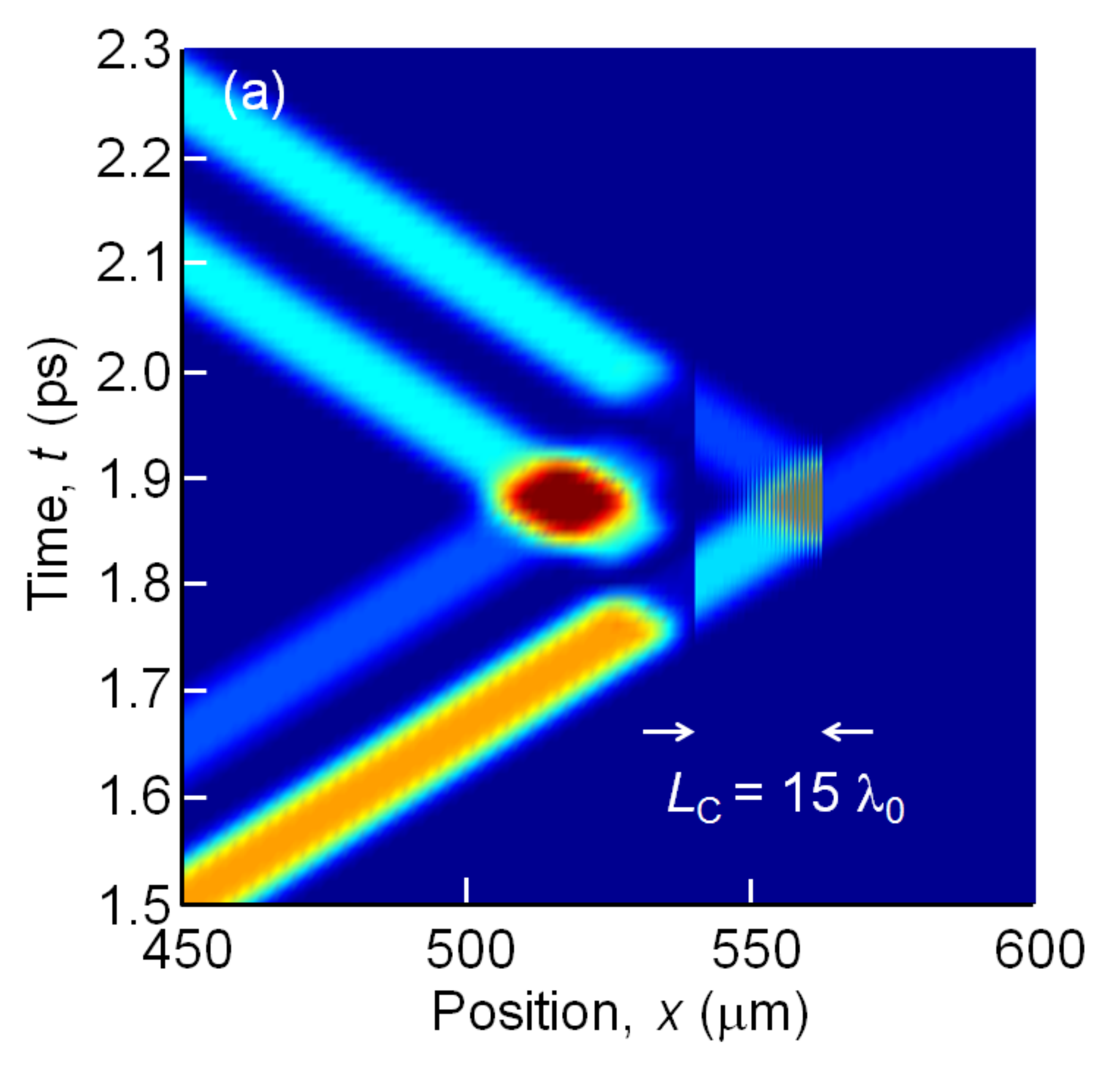} 
\par\end{centering}

\centering{}\includegraphics[width=7.3cm]{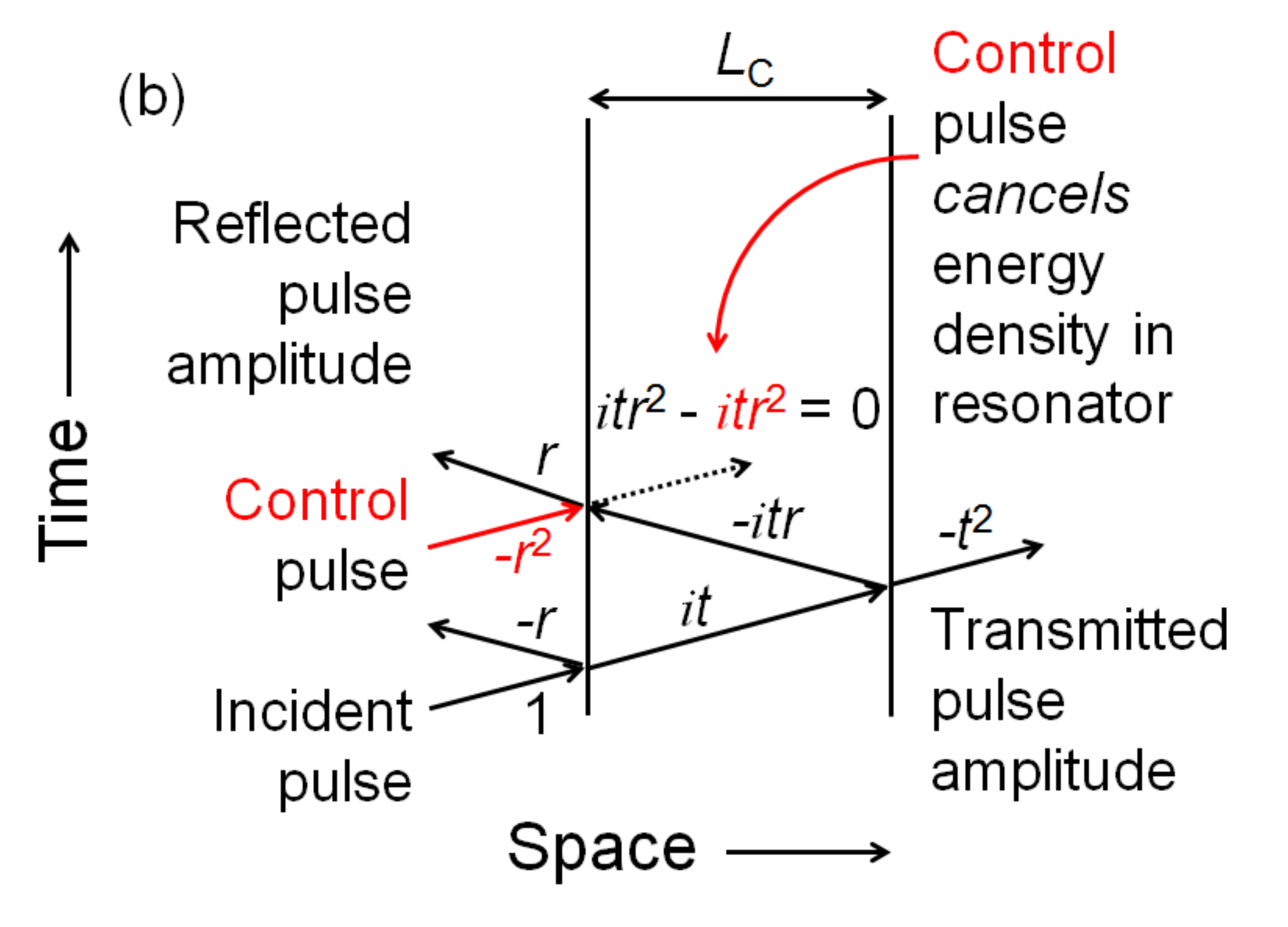} \caption{\label{fig4} {\footnotesize { (a) Space-time photon
energy density plot showing lead and control pulse. The control pulse
eliminates ring-down by removing all photon energy density in the
cavity after exactly one round-trip time, $\tau_{{\rm {RT}}}$. There
is just one transmitted photon pulse. (b) Space-time resonant photon
ray trace showing lead and control amplitudes configured to eliminate
ring-down. Photon pulse parameters are $\hbar\omega_{0}=0.827\ {\rm {eV}}$,
$\hbar\Delta\omega_{{\rm {s}}}=0.207\ {\rm {eV}}$, and $T_{0}\omega_{0}=60$.}}}
\end{figure}

\section{Coherent control of transient response}

\label{sec3} Coherent control of the transient response illustrated
in Fig.~\ref{fig3} may be achieved using photon control pulses.
Similar to Eq.~(\ref{eq2}), the control pulses consist of a coherent
integral of basis functions whose amplitudes $\alpha_{\omega}^{{\rm {cont}}}$
and time delay $t_{\omega}^{{\rm {cont}}}$ are control parameters
that can be optimized. In the following we avoid the use of formal
optimization methods because the geometric series illustrated in Fig.~\ref{fig3}(b)
suggests a simpler intuitive approach.

First we consider a single control pulse that is just an attenuated,
delayed, and phase-shifted version of the lead pulse. Figure~\ref{fig4}(a)
is a space-time photon energy density plot showing a lead pulse and control
pulse initially moving left to right and incident on the Fabry-P/'erot
resonator. In this example the control pulse is configured to eliminate
ring-down after exactly one photon round-trip time in the cavity.
This can be achieved with a control pulse of the same shape that is coherent
with the lead pulse, with resonant amplitude $-r^{2}$ relative to
the lead pulse, and delayed by a time $\tau_{{\rm {RT}}}$. Figure~\ref{fig4}(b)
is a space-time resonant photon ray trace showing lead and control
amplitudes configured to eliminate ring-down.

\begin{figure}[h]
\centering{} \includegraphics[width=4cm]{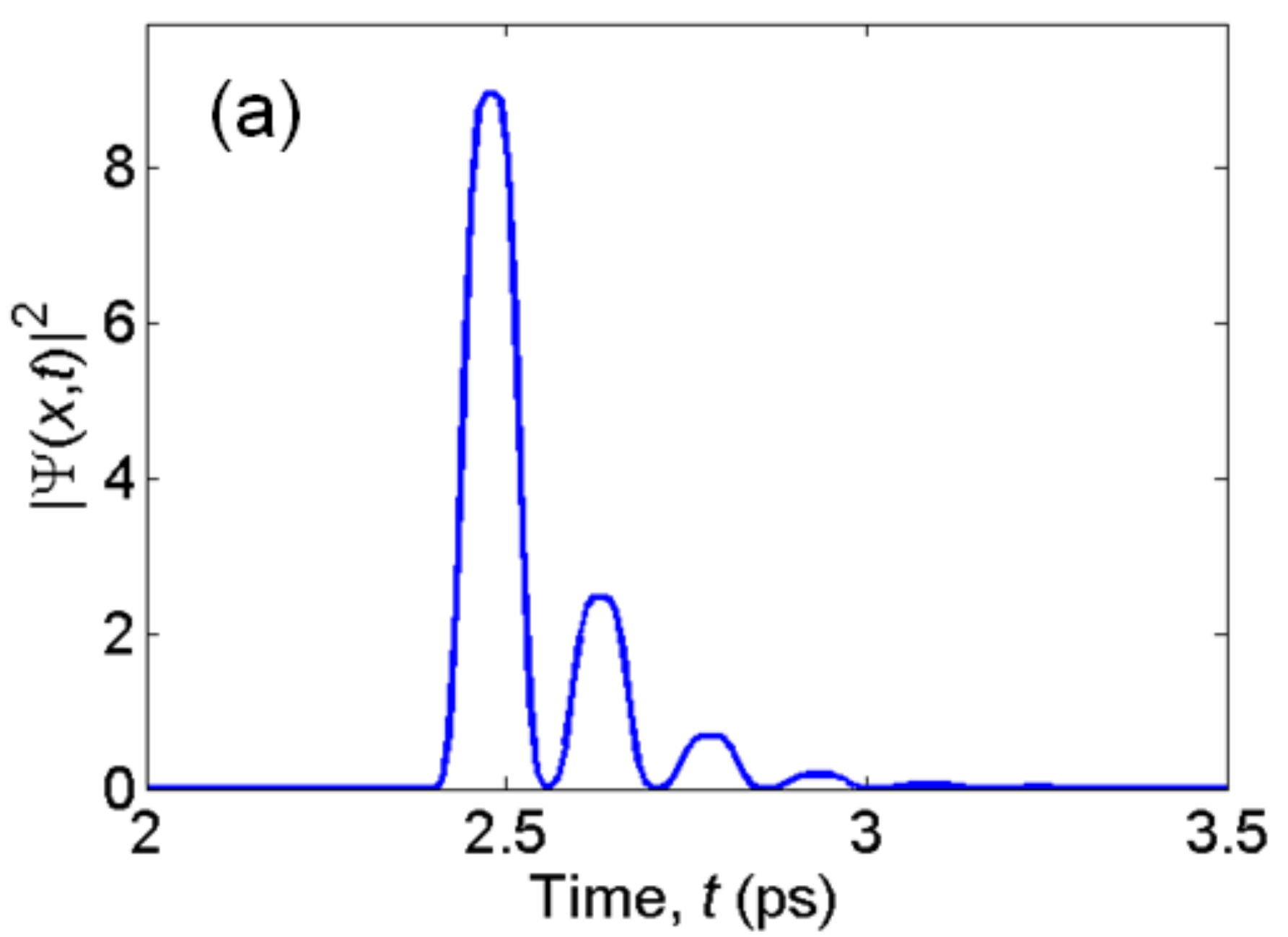} \includegraphics[width=4cm]{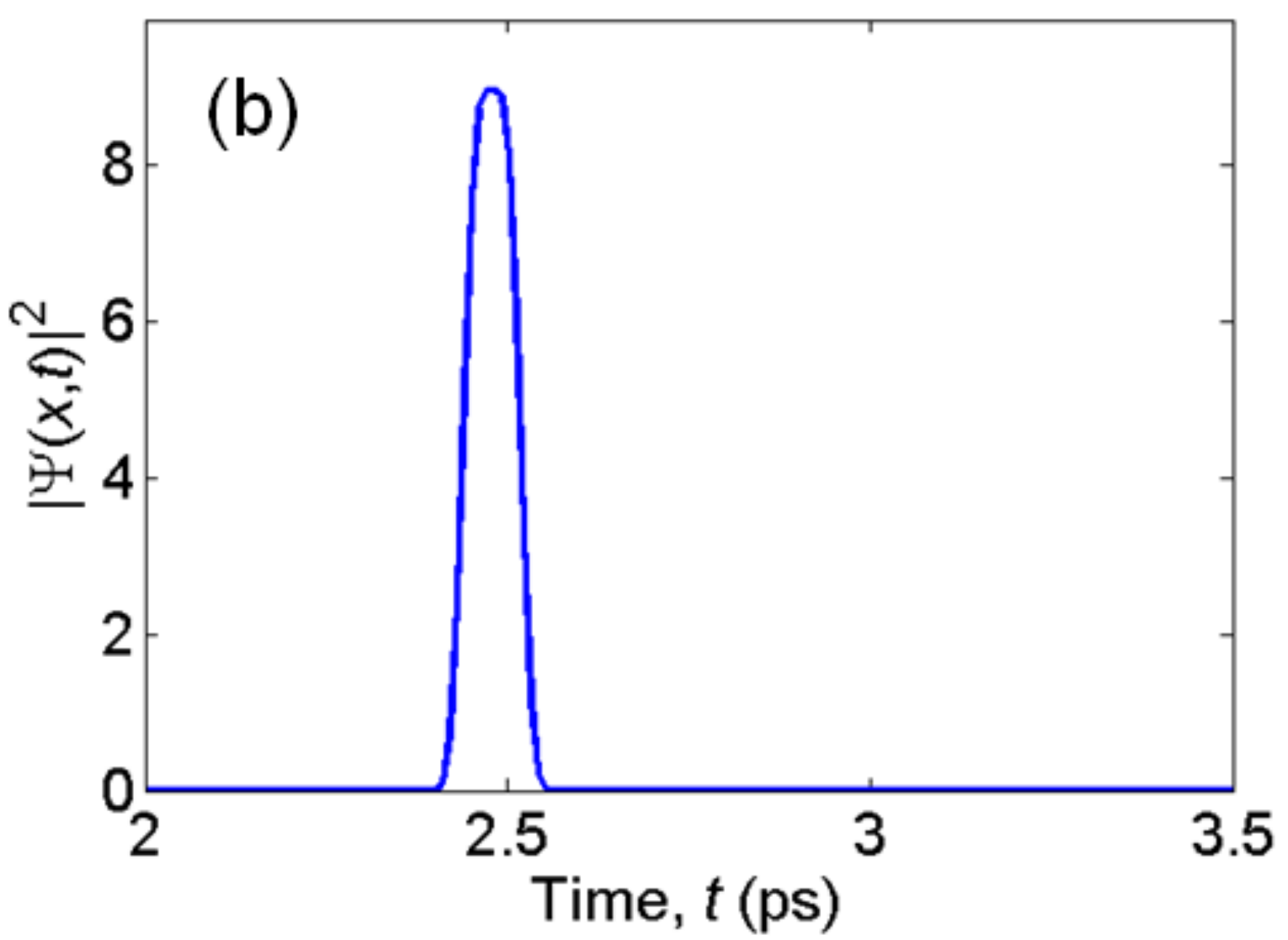}
\caption{\label{fig5}{\footnotesize { (a) Transmitted photon
energy density as a function of time with no control (as in Fig.~\ref{fig3}).
(b) Same as (a) but with a control pulse to eliminate ring-down by removing
all photon energy density in the cavity after exactly one round-trip
time, $\tau_{{\rm {RT}}}$. There is just one transmitted photon pulse.
Photon pulse parameters are $\hbar\omega_{0}=0.827\ {\rm {eV}}$,
$\hbar\Delta\omega_{{\rm {s}}}=0.207\ {\rm {eV}}$, and $T_{0}\omega_{0}=60$.
The photon energy density scale is arbitrary.}}}
\end{figure}

To highlight the difference in the time domain between uncontrolled
ring-down of the resonator and precise control, Fig.~\ref{fig5}
shows the transmitted pulse train for the two situations illustrated
in Figs.~\ref{fig3} and \ref{fig4}. Transmitted photon energy
density as a function of time for the uncontrolled case [Fig.~\ref{fig5}(a)]
consists of a series of pulses whose peaks occur at equally spaced
time intervals $\tau_{{\rm {RT}}}$ and whose peak value decreases
exponentially as $e^{-t/\tau_{{\rm {Q}}}}$. For the controlled case
[Fig.~\ref{fig5}(b)] a coherent control pulse is used to ensure
that there is just one transmitted photon energy density pulse.

\begin{figure}[h]
\begin{centering}
\includegraphics[width=8.5cm]{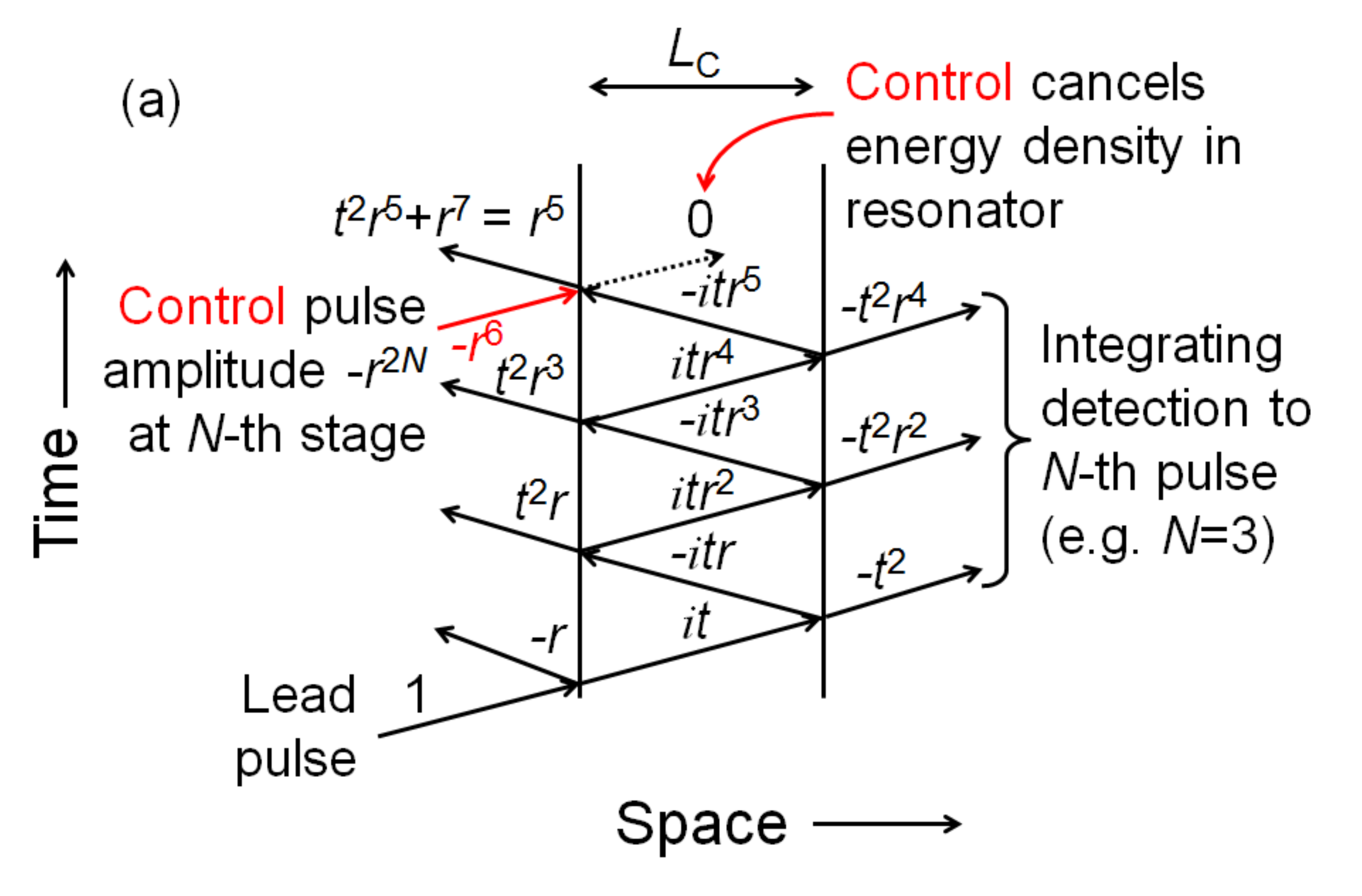} 
\par\end{centering}

\centering{}\includegraphics[width=7.4cm]{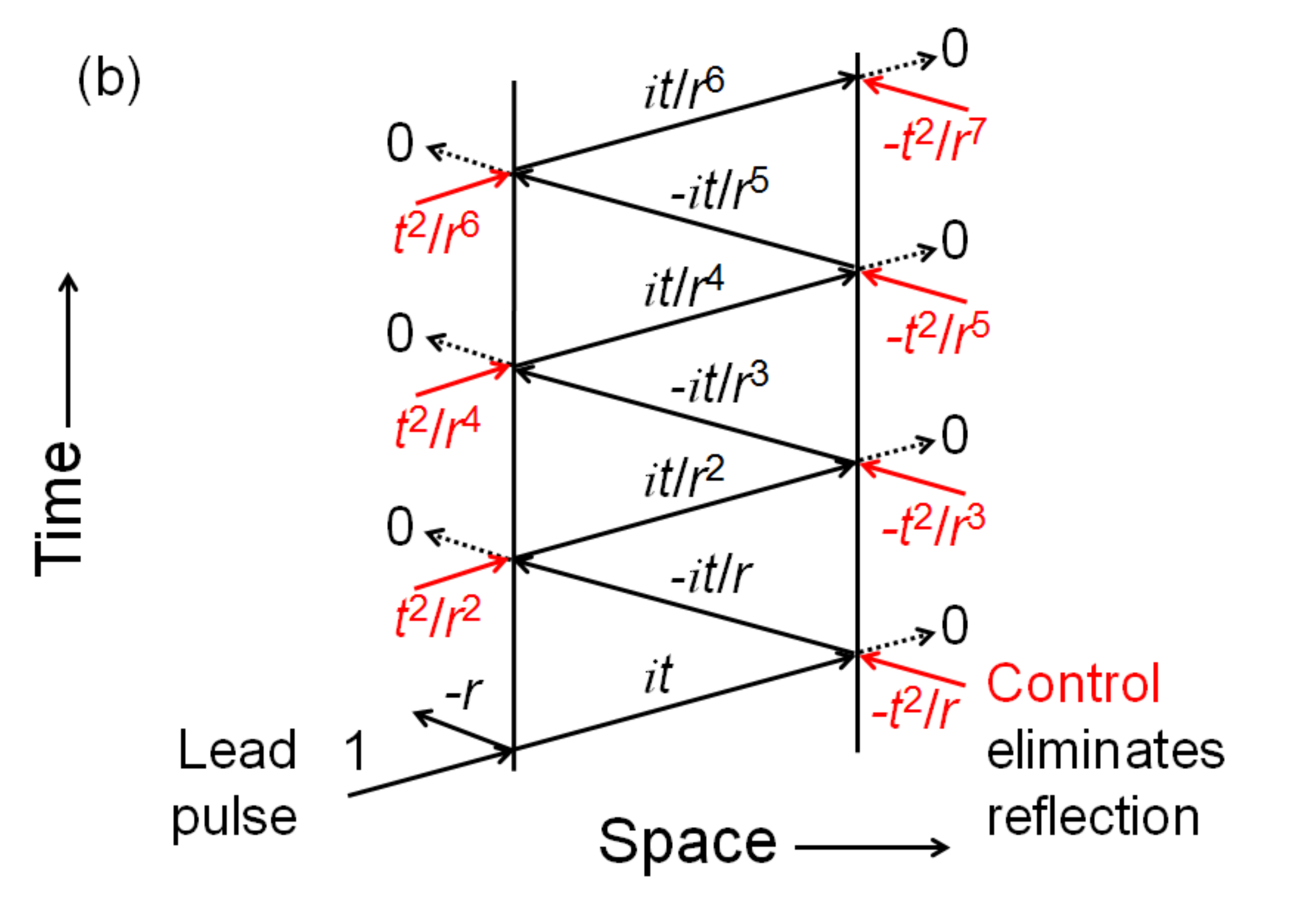} \caption{\label{fig6}(color online) {\footnotesize {(a) Space-time resonant
photon ray trace showing incident lead and control amplitudes configured
to perform a finite geometric sum. (b) Space-time resonant photon
ray trace showing incident lead photon and control amplitudes configured
to create a finite divergent geometric series.}}}
\end{figure}

A coherent control pulse with amplitude $-r^{2N}$ injected at the
$N$-th photon round trip may be used together with an \emph{integrating}
detector to evaluate a finite geometric sum. Figure~\ref{fig6}(a)
illustrates this for the case $N=3$. An integrating photon energy
detector at the output measures this geometric sum as 
\begin{equation}
\left|\sum_{n=0}^{N-1}ax^{n}\right|^{2}=\left|a\frac{1-x^{N}}{1-x}\right|^{2}\ ,\label{eq8}
\end{equation}
 where, on resonance, $x=r^{2}$ and $a=t^{2}$. The sum in Eq.~(\ref{eq8})
is guaranteed to converge in the limit $N\to\infty$ because $|r|<1$.

Figure~\ref{fig6}(b) illustrates that the finite geometric series
in Eq.~(\ref{eq8}) with $|x|>1$, may also be created by using multiple
forward- and reverse-propagating control pulses. In this particular
example coherent photon control pulses are used to \emph{confine}
photon energy density in the resonator. The photon energy density
in the resonator increases according to Eq. (8) because $|r|<1$ and
so $|x|=|e^{i\phi}/r|>1$, where $\phi$ is accumulated phase per
cavity transit.

In general, transient photon dynamics in resonators with input and
output ports may be used to evaluate arbitrary finite sums of the
form 
\begin{equation}
\left|\sum_{n=0}^{N-1}a_{n}x^{n}\right|^{2}\ ,\label{eq9}
\end{equation}
 where complex $a_{n}$ and $x$ are determined by control pulses.

\section{Coherent control of Markovianity}

\label{sec4} So far, we have demonstrated that coherent photon pulses
can control transient photon dynamics in a resonant cavity. Here we
wish to show that such techniques may be understood as controlling
the degree of non-Markovianity exhibited by the system. To demonstrate
control of non-Markovianity in the system it is necessary to adopt
a suitable measure. The definition of a proper measure of non-Markovianity
is currently a topic of debate and various, inequivalent, definitions
have been proposed~\cite{Breuer,non-Markovianity2,non-Markovianity3,non-Markovianity4,non-Markovianity5}.
These definitions suffer from the drawback of being computationally
demanding, and results have been reported only for extremely simple
systems consisting of a single or a few qubits. Recently, a definition
of non-Markovianity valid for Gaussian states (i.e., states satisfying
the Wick theorem) has been proposed which has the advantage of being
computationally tractable even for high-dimensional many-body systems~\cite{Lorenzo}.
In practice one asks if the dynamically evolving Gaussian state under
consideration is consistent with quasifree Markovian dynamics in
the sense of Refs.~\cite{Prosen1,Prosen2,Prosen3,Prosen4}. The answer
is no if the Hilbert-Schmidt distance $D(t):=\left\Vert \Gamma_{1}(t)-\Gamma_{2}(t)\right\Vert _{HS}$
increases for some $t$ for initial states characterized by covariance
matrices $\Gamma_{1,2}(0)$. More details on the precise definition
of $D(t)$ may be found in Ref.~\cite{Lorenzo}. In the following
we are interested in establishing whether the exact evolution occurring
in a spatial subregion $A$ can be considered Markovian according
to the Hilbert-Schmidt distance. To check this we initialize the system
with two different wave-packets $\Psi_{1}(x,0)$ and $\Psi_{2}(x,0)$
which are then evolved according to the exact equation of motion to
$\Psi_{1}(x,t)$ and $\Psi_{2}(x,t)$. The Hilbert-Schmidt measure
$D(t)$ takes the following form~\cite{Lorenzo} 
\begin{align}
D(t) & =\frac{1}{\sqrt{2}}\sqrt{p_{1,1}^{2}+p_{2,2}^{2}-2\left|p_{1,2}\right|^{2}}\label{eq10}\\
p_{i,j} & =\int_{A}\Psi_{i}^{*}(x,t)\Psi_{j}(x,t)dx\,,\label{eq:pij}
\end{align}
where in Eq.~(\ref{eq:pij}) the integral is performed over region
$A$, one of the regions under examination. The system is considered Markovian if $D(t)$
\emph{decreases monotonically} with time for any choice of initial
state. Non-Markovianity is observed when $D(t)$ increases for a pair
of initial states. Instead of considering all possible initial states,
we are interested in quantifying the non-Markovian content of some
physically motivated wave-packets $\Psi_{1}(x,t)$ and $\Psi_{2}(x,t)$.
For simplicity we choose $\Psi_{2}(x,t)=\Psi_{1}(x,t+\tau_{{\rm {M}}})$
for a fixed delay $\tau_{{\rm {M}}}$.

\begin{figure}[h]
\centering{}\includegraphics[width=4cm]{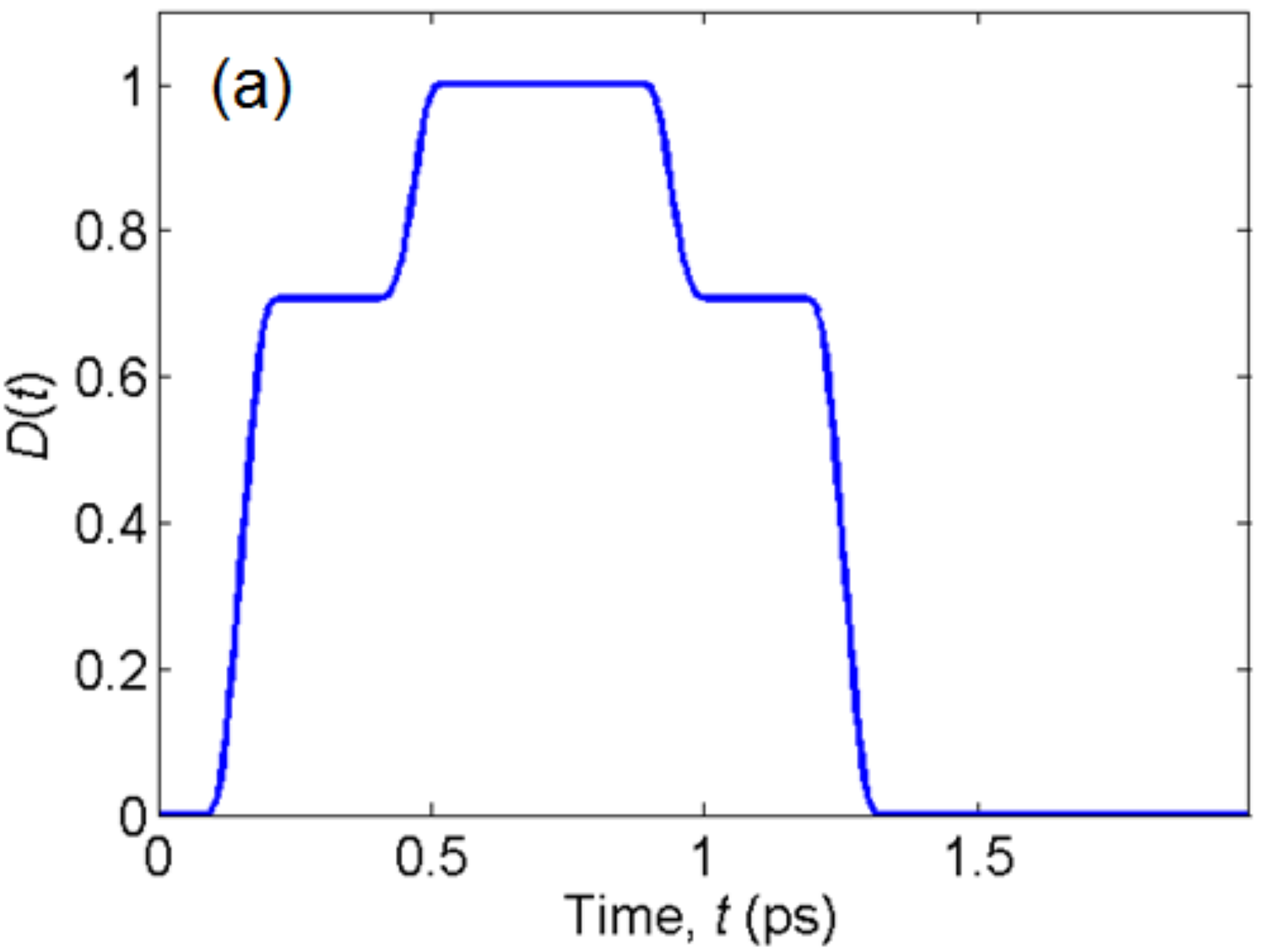}\includegraphics[width=4cm]{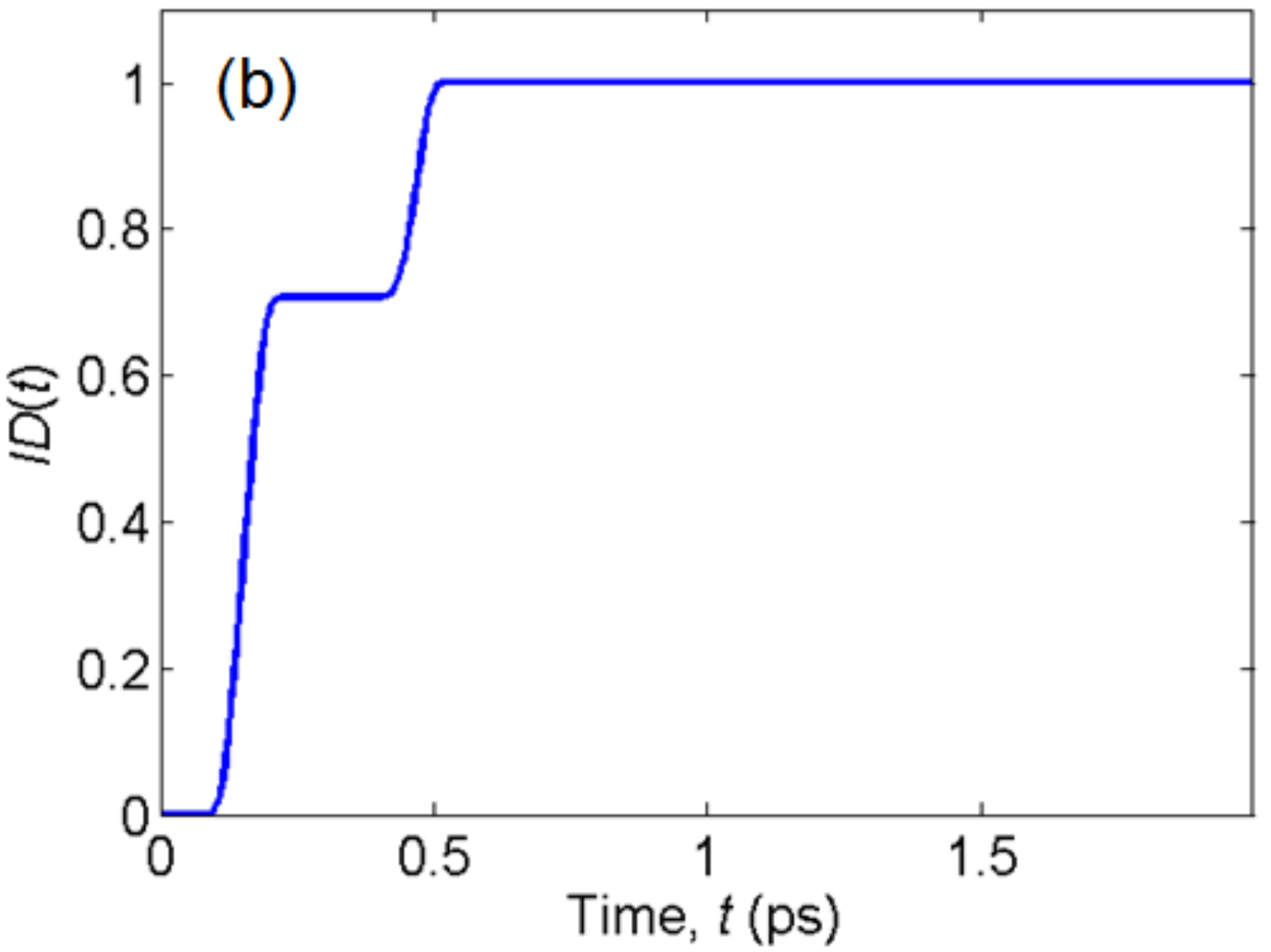}
\caption{\label{fig7} {\footnotesize { (a) Normalized measure
$D(t)$ for rectangular photon pulse $\Psi_{1}(x,t)$ and $\Psi_{2}(x,t)=\Psi_{1}(x,t+\tau_{{\rm {M}}})$
freely propagating through spatial region $A$ of length $L_{{\rm {A}}}=160\times\lambda_{0}$.
When both pulses are simultaneously in spatial region $A$, then $D(t)=1$.
(b) $ID(t)$, the measure of non-Markovian dynamics for the pulse
in (a). Photon pulse parameters are $\hbar\omega_{0}=0.827\ {\rm {eV}}$,
$\hbar\Delta\omega_{{\rm {s}}}=0.207\ {\rm {eV}}$, $T_{0}\omega_{0}=60$,
and $\tau_{{\rm {M}}}/\tau_{0}=60$.}}}
\end{figure}

As an example, consider a rectangular photon pulse propagating in
free space and moving from left to right. It will freely enter, propagate,
and exit the spatial region $A$. Figure~\ref{fig7}(a) shows the
resulting $D(t)$. Initially, both pulses are to the left of region
$A$, so $D(t)=0$. As the first pulse $\Psi_{1}(x,t)$ enters
the region, $D(t)$ increases, eventually reaching a value of $1/\sqrt{2}$.
After time delay $\tau_{{\rm {M}}}$ the second pulse enters and increases
$D(t)$ to its maximum value of unity, corresponding to both pulses
simultaneously being in region $A$. There is a monotonic decrease in $D(t)$
as $\Psi_{1}(x,t)$ and $\Psi_{2}(x,t)$ leave and information leaks
out of region $A$, indicating pure Markovian behavior.

A direct measure of non-Markovian dynamics is simply the positive
contributions to $D(t)$. Defining $\sigma(t)=dD(t)/dt$, the total
non-Markovian content after a time $t$ is 
\begin{equation}
ID(t)=\int\displaylimits_{\sigma(\tau)>0\cap[0,t]}\,\sigma(\tau)d\tau\ .\label{eq11}
\end{equation}
 The total non-Markovianity of the trajectories under consideration
is $ID:=ID(\infty)$, where a larger value of $ID$ means more non-Markovian.
Figure~\ref{fig7}(b) shows the result of calculating $ID(t)$ for
the non-interacting rectangular photon pulse considered in (a). As
expected, after the two freely propagating pulses enter region $A$
there is no further increase in $ID(t)$. The purely Markovian dynamics
of a system with no scattering results in a constant value for $ID(t)$.

\begin{figure}[h]
\centering{} \includegraphics[width=85mm]{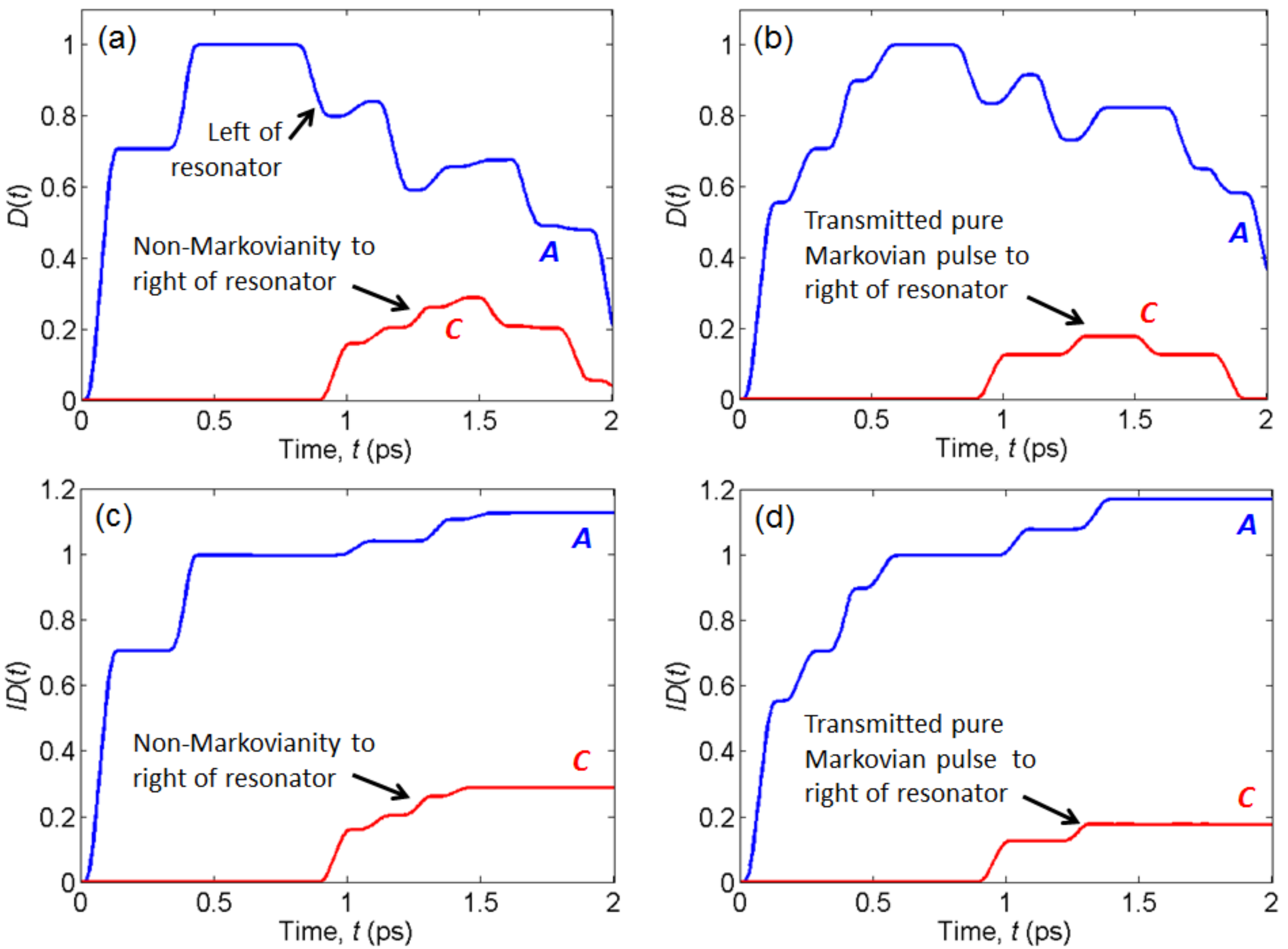} \caption{\label{fig8}{\footnotesize { (a) $D(t)$ for rectangular
photon pulse $\Psi_{1}(x,t)$ and $\Psi_{2}(x,t)=\Psi_{1}(x,t+\tau_{{\rm {M}}})$
in spatial region $A$ [blue] to the left of the dielectric resonator
and region $C$ [red] to the right of the resonator. $L_{{\rm {A}}}=160\times\lambda_{0}$
and $L_{{\rm {C}}}=120\times\lambda_{0}$. (c) $ID(t)$ for uncontrolled
ring-down of $D(t)$ shown in (a). The calculated duration of non-Markovianity
in region $A$ to the left of the dielectric resonator is limited
by the pulse leaving the domain. (b) $D(t)$ and (d) $ID(t)$ illustrate the 
use of a single control pulse to reduce non-Markovianity in transient
dynamics by removing all photon energy in the resonator after exactly
one cavity round-trip time, $\tau_{{\rm {RT}}}$. Photon pulse parameters
are $\hbar\omega_{0}=0.827\ {\rm {eV}}$, $\hbar\Delta\omega_{{\rm {s}}}=0.207\ {\rm {eV}}$,
$T_{0}\omega_{0}=60$, and $\tau_{{\rm {M}}}/\tau_{0}=60$.}}}
\end{figure}

Photon dynamics are very different in the presence of a Fabry-P/'erot
resonator because energy density can be scattered and stored. Figure~\ref{fig8}
shows the result of calculating $D(t)$ and $ID(t)$ for a rectangular
photon pulse initially incident from the left in spatial regions as
defined in Figure~\ref{fig1}(b). We ask whether the dynamics in
finite-sized regions $A$ and $C$ can be considered Markovian and
try to quantify its non-Markovianity content. Figures~\ref{fig8}(a)
and \ref{fig8}(c) show results for $D(t)$ and $ID(t)$ for a single initial
rectangular pulse, while Figs.\ref{fig8}(b) and \ref{fig8}(d) refer to the dynamics with a
control pulse applied to remove \emph{all} photon energy density inside
the resonator after just one photon round-trip time $\tau_{{\rm {RT}}}$,
as shown in Fig.~\ref{fig4}(b).

The functions $D(t)$ and $ID(t)$ display complex temporal patterns,
some features of which can readily be connected with the physics of
the resonator. For example, consider the curve labelled $C$ in Fig.~\ref{fig8}(c).
Here we sent a control pulse whose effect is to eliminate completely
the ring-down structure at the output (right) of the resonator. Consequently
the single transmitted pulse has a behavior identical to that for the
simple dynamics illustrated in Fig.~\ref{fig7}.

However, a simple picture seems to emerge from Fig.~\ref{fig8}, namely
the \emph{total} non-Markovian content of the dynamics restricted
to region $C$ (right of the resonator) diminishes after application
of the control pulse. On the other hand, the total non-Markovian content
restricted to region $A$ (left of the resonator) increases in the
presence of the control pulse. The sum of total non-Markovianity in
regions $A$ and $C$ is greater for the uncontrolled case than for the controlled case.

\begin{figure}[t]
\centering{} \includegraphics[width=0.7\columnwidth]{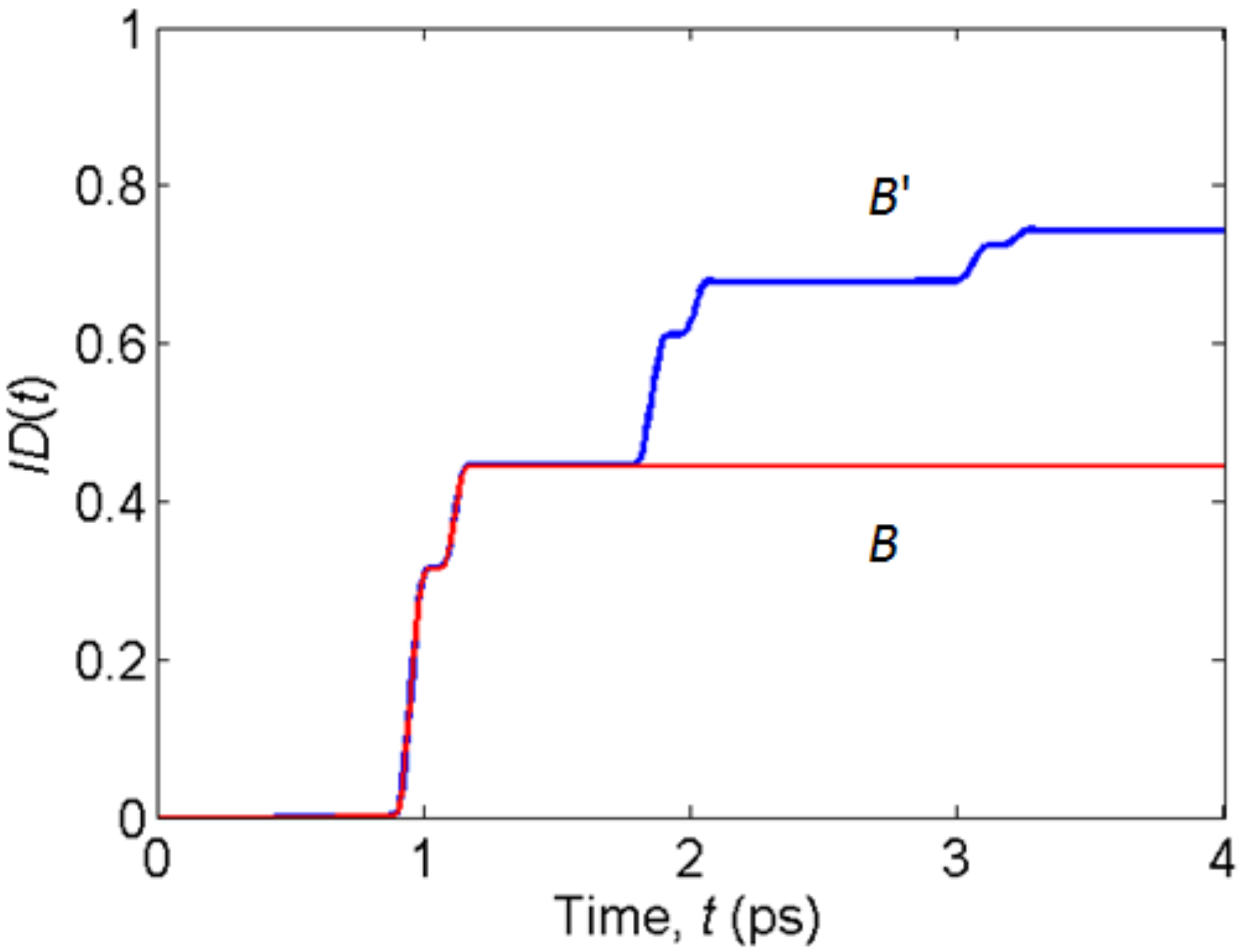} \caption{\label{fig9}{\footnotesize { $ID(t)$ for a resonator
with cavity length $L_{{\rm {B}}}=120\times\lambda_{0}$, which defines
region $B$, and a half space in the resonator cavity of length $L_{{\rm {B}}}/2$
adjacent to the left mirror, which defines region $B'$. In the absence
of a coherent control pulse, $ID(t)$ is purely Markovian for subspace
$B$ and is non-Markovian for subspace $B'$. Photon pulse parameters
are $\hbar\omega_{0}=0.827\ {\rm {eV}}$, $\hbar\Delta\omega_{{\rm {s}}}=0.207\ {\rm {eV}}$,
$T_{0}\omega_{0}=60$, and $\tau_{{\rm {M}}}/\tau_{0}=30$.}}}
\end{figure}

Non-Markovianity depends on subspace size and the location to which it
is referred, so this naturally provides another approach to its
control. For example, evaluating $D(t)$ in region $B$ inside the
resonator involves spatial integrals over the entire cavity length,
$L_{{\rm {B}}}$. However, if one evaluates $D(t)$ in a smaller portion
of the cavity, then information flows in and out of that subspace as
energy density builds up or decays in the resonator. Oscillatory values
of $D(t)$ on a rising or falling background result, indicating non-Markovian
contributions in a small subspace inside the resonator. These oscillations
are averaged out when the subspace is increased to include the complete
resonator cavity of length $L_{{\rm {B}}}$ (region $B$). This illustrates
the fact that small subspaces can be \emph{tuned} to exhibit \emph{enhanced}
non-Markovian effects.

To show how spatial placement of subspace determines Markovianity,
consider a resonator with cavity length $L_{{\rm {B}}}=120\times\lambda_{0}$
that defines region $B$ and a half space in the resonator cavity
of length $L_{{\rm {B}}}/2$ adjacent to the left mirror, which defines
region $B'$. As shown in Fig.~\ref{fig9}, in the absence of coherent
control, transient photon evolution dynamics is more non-Markovian
for the half-cavity subspace $B'$ than for the full-cavity subspace
$B$. Placing the half-cavity subspace $B'$ symmetrically about the
center of the resonator does not remove non-Markovian dynamics. This
is because resonator energy density stored outside $B'$ is reflected
by the mirrors back into $B'$, causing the non-Markovian behavior.

\section{Conclusions}

\label{sec5} We have applied a Hilbert-Schmidt measure of non-Markovianity
to a photon interacting with a symmetric lossless dielectric resonator.
Non-Markovian transient photon dynamics in a resonator subsystem coupled
to continuum states is shown to be controlled using coherent pulses.
Transient photon dynamics can be controlled to within a photon resonator
transit time. The underlying physical mechanisms used to control the dynamics at
resonance are conveniently described using interference arising in
finite geometric series with complex amplitudes. In general, coherent
pulses, combined with a suitable choice of spatial subspace, may be
used to both create and control a wide range of non-Markovian transient
dynamics in photon-resonator systems. This initial study has revealed
a richness in both the physics and control of single-photon transient
dynamics interacting with a resonator, suggesting further study is
warranted.

\noindent \acknowledgments \indent This research is partially supported
by ARO MURI Grant No. W911NF-11-1-0268.

\appendix
%dummy comment inserted by tex2lyx to ensure that this paragraph is not empty
%dummy comment inserted by tex2lyx to ensure that this paragraph is not empty
%dummy comment inserted by tex2lyx to ensure that this paragraph is not empty
%dummy comment inserted by tex2lyx to ensure that this paragraph is not empty
%dummy comment inserted by tex2lyx to ensure that this paragraph is not empty
\numberwithin{equation}{section}

\section{The photon wave function}

\label{sec:Photon}

\subsection{Introduction}

To justify the use of the single photon wave function we present a
simplified version of the arguments given in Refs.~\cite{Birula,Sipe,Smith,Abram}.
This is done by first quantizing the electromagnetic field and restricting
the discussion to wave propagation in one spatial dimension and lossless
dielectrics. The corresponding quantized energy density operator in
the dielectric is not diagonal when expressed in terms of the free-field
operators. However, the energy density may be diagonalized through
a unitary Bogolyubov transformation that relates the dielectric creation
and annihilation operators to the free-field creation and annihilation
operators. The abrupt perturbation at the air-dielectric interface
may be viewed as projecting free-waves onto refracted waves using
the ``sudden approximation.'' Continuity and smoothness are guaranteed
via the field interface conditions.

\subsection{The single-photon wave function in vacuum from the quantized electromagnetic
field}

The electromagnetic field in vacuum may be quantized in the Coulomb gauge
to give: 
\begin{subequations}
\begin{align}
\vec{A}(x^{\mu}) & =\left(\frac{\hbar}{\epsilon_{0}}\right)^{1/2}\int\frac{d^{3}k}{(2\pi)^{3/2}}\frac{1}{\sqrt{2\omega}}\nonumber \\
 &\times \sum_{r=1}^{2}\left(\vec{\epsilon}_{r}(\vec{k})a_{\vec{k},r}e^{-ik_{\mu}x^{\mu}}+\vec{\epsilon}_{r}(\vec{k})^{\ast}a_{\vec{k},r}^{\dagger}e^{ik_{\mu}x^{\mu}}\right)\ ,\\
\vec{E}(x^{\mu}) & =\left(\frac{\hbar}{\epsilon_{0}}\right)^{1/2}i\int\frac{d^{3}k}{(2\pi)^{3/2}}\sqrt{\frac{\omega}{2}}\nonumber \\
 & \times\sum_{r=1}^{2}\left(\vec{\epsilon}_{r}(\vec{k})a_{\vec{k},r}e^{-ik_{\mu}x^{\mu}}-\vec{\epsilon}_{r}(\vec{k})^{\ast}a_{\vec{k},r}^{\dagger}e^{ik_{\mu}x^{\mu}}\right)\ ,\\
\vec{B}(x^{\mu}) & =\left(\frac{\hbar}{\epsilon_{0}}\right)^{1/2}i\int\frac{d^{3}k}{(2\pi)^{3/2}}\frac{1}{\sqrt{2\omega}}\nonumber \\
 &\times \sum_{r=1}^{2}\left(\left(\vec{k}\times\vec{\epsilon}_{r}(\vec{k})\right)a_{\vec{k},r}e^{-ik_{\mu}x^{\mu}}\right.\nonumber \\
 & \left.-\left(\vec{k}\times\vec{\epsilon}_{r}(\vec{k})^{\ast}\right)a_{\vec{k},r}^{\dagger}e^{ik_{\mu}x^{\mu}}\right),
\end{align}
\end{subequations}
\noindent where $x^{\mu}=(ct,\vec{x})$, $k^{\mu}=(\omega/c,\vec{k})$,
$\vec{\epsilon}_{r}$ is a polarization vector satisfying $\vec{k}\cdot\vec{\epsilon}_{r}(\vec{k})=0$, and
$(a_{\vec{k},r},a_{\vec{k},r}^{\dagger})$ are annihilation and creation
operators of a single plane wave excitation with momentum $\vec{k}$
and polarization $r$: 
\begin{equation}
\left[a_{\vec{k},r},a_{\vec{k}',s}^{\dagger}\right]=\delta_{rs}\delta^{(3)}(\vec{k}-\vec{k}')\ .
\end{equation}
 The (normal ordered) Hamiltonian and momentum are, respectively, given by
\begin{align}
:\! H\!: & =\hbar\int d^{3}k\ \omega_{k}\sum_{r}a_{\vec{k},r}^{\dagger}a_{\vec{k},r}\ ,\\
:\!\vec{P}\!: & =\hbar\int d^{3}k\ \vec{k}\sum_{r}a_{\vec{k},r}^{\dagger}a_{\vec{k},r}\ .
\end{align}
 The vacuum $\ket{0}$ is defined by requiring $a_{k}\ket{0}=0$ for
all $k$. The one-particle state 
\begin{equation}
\ket{\vec{k},r}\equiv\sqrt{\frac{\omega_{k}}{c}}a_{\vec{k},r}^{\dagger}\ket{0}\ ,
\end{equation}
 is an eigenstate of the momentum operator with momentum $\vec{k}$
and polarization $r$. The reason for the $\sqrt{\omega_{k}}$ prefactor
is that this makes the orthogonality condition given by 
\begin{equation}
\langle\vec{k},r|\vec{k}',r'\rangle=\frac{\omega_{k}}{c}\delta^{(3)}(\vec{k}-\vec{k}')\delta_{rr'}\ ,
\end{equation}
 Lorentz invariant~\cite{Peskin}. The one-particle completeness condition
is then given by: 
\begin{equation}
\ident_{\mathrm{1-particle}}=\sum_{r}\int d^{3}k\frac{c}{\omega_{k}}\ketbra{\vec{k},r}{\vec{k},r}\ .\label{eqt:kcomplete}
\end{equation}
 For a momentum state with $\vec{k}=k\hat{e}_{z}$, we can choose
$\vec{\epsilon}_{1}\equiv\vec{\epsilon}_{+}=-(\hat{e}_{x}+i\hat{e}_{y})/\sqrt{2}$,
$\vec{\epsilon}_{2}\equiv\vec{\epsilon}_{-}=(\hat{e}_{x}-i\hat{e}_{y})/\sqrt{2}$,
such that the momentum state satisfies
\begin{align}
:\! H\!:\ \ket{k\hat{e}_{z},\pm} & =\hbar ck\ket{k\hat{e}_{z},\pm}\ ,\\
:\!\vec{P}\!:\ \ket{k\hat{e}_{z},\pm} & =\hbar k\hat{e}_{z}\ket{k\hat{e}_{z},\pm}\ ,\\
S_{z}\ket{k\hat{e}_{z},\pm} & =\pm\ket{k\hat{e}_{z},\pm}\ ,
\end{align}
 where $S_{z}=-i\hbar\left(\hat{e}_{x}\otimes\hat{e}_{y}-\hat{e}_{y}\otimes\hat{e}_{x}\right)$
is the $z$-component of the angular momentum operator. Therefore,
let us consider a single particle state $\ket{\psi}$. We define the
(vector) momentum space wave function of helicity $r$: 
\begin{equation}
\vec{\psi}_{r}(\vec{k})=\braket{\vec{k},r}{\psi}\ .
\end{equation}
 This is a vector due to the spin-1 nature of the photon and how the
state must behave under the angular momentum operator. Furthermore,
it satisfies
\begin{equation}
\vec{k}\cdot\vec{\psi}_{r}(\vec{k})=0\ .\label{eqt:transverse}
\end{equation}
 The normalization of this momentum-space wave function is determined
by the completeness condition: 
\begin{equation}
1=\braket{\psi}{\psi}=\sum_{r}\int d^{3}k\frac{c}{\omega_{k}}\vec{\psi}_{r}(\vec{k})^{\dagger}\vec{\psi}_{r}(\vec{k})\ .
\end{equation}
 This normalization of the momentum-space wave function matches that
defined by Ref.~\cite{Birula}. We define the position-space wave
function $\vec{\phi}(x)$ of the state to be simply the Fourier transform
of $\vec{\psi}(k)$: 
\begin{equation}
\vec{\phi}_{r}(\vec{x})=\int\frac{d^{3}k}{(2\pi)^{3/2}}e^{i\vec{k}\cdot\vec{x}}\vec{\psi}_{r}(\vec{k})\ .
\end{equation}
 This implies that
\begin{align}
\bra{\psi}:\! H\!:\ket{\psi} & =\sum_{r}\int d^{3}x\ \vec{\phi}_{r}(\vec{x})^{\dagger}\vec{\phi}_{r}(\vec{x})\nonumber \\
 & =\sum_{r}\int d^{3}k\ \vec{\psi}_{r}(\vec{k})^{\dagger}\vec{\psi}_{r}(\vec{k})\,,\label{eqt:EnergyDensity}
\end{align}
 Furthermore, Eq.~\eqref{eqt:EnergyDensity} suggests that we
interpret $\vec{\psi}_{r}(\vec{k})^{\dagger}\vec{\psi}_{r}(\vec{k})d^{3}k$
as the energy density in the shell $\vec{k}$ and $\vec{k}+d\vec{k}$
in momentum space rather than a probability density~\cite{Smith}.
Finally, we note that
\begin{align}
\bra{\vec{k},\sigma}:\! H\!:\ket{\psi} & =c\hbar|\vec{k}|\vec{\psi}_{\sigma}(\vec{k})\nonumber \\
 & =c\hbar\sigma\left(\vec{s}\cdot\vec{k}\right)\vec{\psi}_{\sigma}(\vec{k})\nonumber \\
 & =ic\hbar\sigma\vec{k}\times\vec{\psi}_{\sigma}(\vec{k})\,,
\end{align}
 where $\vec{s}=(s_{x},s_{y},s_{z})$ are the three spin-1 matrices
(generators of rotations for spin 1 particles; angular momentum is
$\vec{S}=\hbar\vec{s}$), and we use the feature of spin-1 matrices
that $\vec{a}\times\vec{b}=-i(\vec{a}\cdot\vec{s})\vec{b}$~\cite{Smith}.
Since the Hamiltonian is the generator of time translations, we have
our Schrödinger equation: 
\begin{equation}
i\hbar\partial_{t}\vec{\psi}_{\sigma}(\vec{k},t)=ic\hbar\sigma\vec{k}\times\vec{\psi}_{\sigma}(\vec{k},t)\ ,
\end{equation}
 or in position space,
\begin{equation}
i\hbar\partial_{t}\vec{\phi}_{\sigma}(\vec{x},t)=c\hbar\sigma\nabla\times\vec{\phi}_{\sigma}(\vec{x},t)\ .
\end{equation}
 Applying another $i\hbar\partial_{t}$, we recover the Helmholtz
equation: 
\begin{equation}
-\partial_{t}^{2}\vec{\phi}_{\sigma}(\vec{x},t)=c\nabla\times\left(c\nabla\times\vec{\phi}_{\sigma}(\vec{x},t)\right)=-c^{2}\nabla^{2}\vec{\phi}_{\sigma}(\vec{x},t)\ ,
\end{equation}
 where in the last equality we used Eq.~\eqref{eqt:transverse}.
%%%%%%%%%%%%%%%%%%%%%%%%%%%%%%%%%

\subsection{Quantization in a linear lossless dielectric}

%%%%%%%%%%%%%%%%%%%%%%%%%%%%%%%%%
One way to study the effect of the presence of the lossless linear
medium is to consider the modified Hamiltonian density \cite{Abram}
\begin{equation}
\mathcal{H}=\frac{1}{2}\left(\epsilon_{0}\vec{E}^{2}+\frac{1}{\mu_{0}}\vec{B}^{2}+\chi\vec{E}^{2}\right)\ .
\end{equation}
 (For other methods see Ref.~\cite{Glauber}.) Inserting the quantized
fields in vacuum into this expression clearly shows that the Hamiltonian
density operator is not diagonal in terms of the free-field creation
and annihilation operators $(a_{\vec{k},\sigma}^{\dagger},a_{\vec{k},\sigma})$.
Nevertheless, the Hamiltonian density can be diagonalized in terms
of ``refracted-wave'' operators $(b_{\vec{k},\sigma}^{\dagger},b_{\vec{k},\sigma})$
via a Bogolyubov transformation~\cite{Abram}. In this basis, the
Hamiltonian density has the same eigenvalues as in vacuum, but the
momentum operator is renormalized by a factor of the index of refraction,
such that the results match the known results from classical optics.

At an abrupt vacuum-dielectric interface, where the permittivity changes
from $\epsilon=\epsilon_{0}$ to $\epsilon=\sqrt{n_{{\rm {r}}}}\epsilon_{0}$,
we can treat the change in the momentum operator (from the vacuum
form to the renormalized form) in the ``sudden approximation''~\cite{Abram}.
For example, a single excitation $a_{\vec{k},\sigma}^{\dagger,\sigma}\ket{0}$
gets projected to~\cite{Abram}
\begin{equation}
\left(\frac{2\sqrt{n_{{\rm {r}}}}}{n_{{\rm {r}}}+1}b_{\vec{k},\sigma}^{\dagger}+\frac{n_{{\rm {r}}}-1}{n_{{\rm {r}}}+1}a_{\vec{k},\sigma}^{\dagger}\right)\ket{0}\ ,
\end{equation}
 such that the probability of reflection and transmission agrees with
the classical result for energy reflection and transmission. With
these results in mind, we can generalize our equation for the single
photon wave function to satisfy the Helmholtz equation in the presence
of a lossless dielectric: 
\begin{equation}
\nabla\times\left(\frac{1}{\mu(x)}\nabla\times\vec{\psi}_{\sigma}(\vec{x},t)\right)=-\epsilon(x)\partial_{t}^{2}\vec{\psi}_{\sigma}(\vec{x},t)\ .
\end{equation}
 For a linearly polarized, transverse field which propagates
in the $x$ direction one has $\vec{\psi}_{\sigma}(\vec{x},t)=(0,\phi_{\sigma}(x,t),0)$,
and one  recovers Eq.~(\ref{eq3}) after a time Fourier
transform. 
%\bibliography{ref}


\end{document}